\documentclass[CRPHYS,Unicode,manuscript]{cedram}

\usepackage{glossaries}
 
\makeglossaries

\usepackage{amssymb}

\title{Loop currents in quantum matter}

\author{\firstname{Philippe} \lastname{Bourges}} 
\address{Universit\'e Paris-Saclay, CNRS, CEA, Laboratoire L{\'e}on Brillouin,  91191, Gif-sur-Yvette, France}
\email[P. Bourges]{philippe.bourges@cea.fr}
\email[D. Bounoua]{dalila.bounoua@cea.fr}
\email[Y. Sidis]{yvan.sidis@cea.fr}

\thanks{We acknowledge financial supports from the project NirvAna (Contract ANR-14-OHRI-0010) of the Agence National de la Recherche (ANR) French 
agency.} 

\author{\firstname{Dalila} \lastname{Bounoua}}
\addressSameAs{1}{Universit\'e Paris-Saclay, CNRS, CEA, Laboratoire L{\'e}on Brillouin,  91191, Gif-sur-Yvette, France}

\author{\firstname{Yvan} \lastname{Sidis}}
\addressSameAs{1}{Universit\'e Paris-Saclay, CNRS, CEA, Laboratoire L{\'e}on Brillouin,  91191, Gif-sur-Yvette, France}

\thanks{\   \\  
 \it Prix CEA "science et innovation" de l’Académie des sciences 2020}


\begin{abstract} 

In many quantum materials, strong electron correlations lead to the emergence of  new states of matter. In particular, the study in the last decades of the complex phase diagram of high temperature superconducting cuprates  highlighted  intra-unit-cell electronic instabilities breaking discrete Ising-like symmetries, while preserving  the lattice translation invariance. Polarized neutron diffraction experiments have provided compelling evidences supporting a new form of  intra-unit-cell magnetism, emerging 
concomitantly with the  so-called pseudogap state of these materials. This observation is currently  interpreted as the magnetic hallmark of an intra-unit-cell  loop current order, breaking both parity and time-reversal 
symmetries. More generally, this magneto-electric state is likely to exist in a wider class of quantum materials beyond superconducting cuprates. For instance,  it has  been already observed in hole-doped Mott insulating iridates or in the spin liquid state of hole-doped 2-leg ladder cuprates.

\end{abstract}

\begin{document}

\newglossaryentry{af}{name=AF, description={antiferromagnetic/antiferromagnetism}}
\newglossaryentry{arpes}{name=ARPES,  description={angle resolved photoemission}}
\newglossaryentry{cdw}{name=CDW, description={charge density wave}}
\newglossaryentry{iuc}{name=IUC, description={intra-unit-cell}}
\newglossaryentry{lc}{name=LC, description={loop current}}
\newglossaryentry{lt}{name=LT, description={lattice translation}}
\newglossaryentry{musr}{name=\gls{musr}, description={muon spectroscopy/muon-spin rotation}}
\newglossaryentry{nmr}{name=NMR, description={nuclear magnetic resonance}}
\newglossaryentry{parity}{name=P, description={parity}}
\newglossaryentry{pdw}{name=PDW, description={pair density wave}}
\newglossaryentry{pg}{name=PG, description={pseudogap}}
\newglossaryentry{pnd}{name=PND, description={polarized neutron scattering}}
\newglossaryentry{rotation}{name=R, description={rotational}}
\newglossaryentry{shg}{name=SHG, description={second harmonic generation}}
\newglossaryentry{stm}{name=STM, description={scanning tunneling microscopy}}
\newglossaryentry{time}{name=T, description={time}}

\maketitle


 \selectlanguage{french}
 \section*{Résumé en français}
Dans de nombreux matériaux quantiques, de fortes corrélations électroniques conduisent à l'émergence de nouveaux états 
de la matière. En particulier, l'étude du diagramme de phase complexe des cuprates supraconducteurs à haute température critique au cours des dernières décennies a mis en évidence des instabilités électroniques à l'intérieur de la maille 
atomique élémentaire brisant des symétries discrètes de type Ising, tout en préservant l'invariance par translation du réseau. Des expériences de diffraction des neutrons polarisés ont fourni des preuves convaincantes en faveur d'une nouvelle forme de magnétisme  émergeant en même temps que l'état dit de "pseudo-gap" de ces matériaux. Cette observation est actuellement interprétée comme étant la signature magnétique d'un ordre  de boucles de courants  à l'intérieur de la maille élémentaire, brisant à la fois les symétries de parité et de renversement du temps. Plus généralement, cet état magnéto-électrique est susceptible d'exister dans une classe plus large de matériaux quantiques au-delà des cuprates supraconducteurs. Par exemple, il a déja été observé dans les composés à 
base d'iridium qui sont des isolants de Mott dopés en trous ou bien dans l'état liquide de spin des cuprates à échelles de spins à deux montants dopés en trous.

\selectlanguage{english}

\section{Introduction}

Strong electron correlations in condensed matter lead to the emergence of 
novel phenomena and exotic states. None of these collective phenomena is encompassed by the simple addition of individual electrons but  results instead from many-body effects as it was advocated by P. W. Anderson almost 50 years ago \cite{Anderson72}.  Spin liquids, Mott physics, strange metals, high-temperature superconductivity and its related pseudogap phase are all typical examples of these novel states of matter, generically labeled as quantum matter. 

The study of the complex phase diagram of hole-doped superconducting cuprates highlighted the existence of a mysterious state of matter, the so-called pseudogap (\gls{pg}) metal  \cite{Keimer15,Proust19,Varma20}, where large portions of the Fermi surface are gapped out and  only Fermi arcs survive. The \gls{pg} state is characterized by a significant loss of electronic entropy \cite{Loram01} below a certain temperature, so-called T$^*$,  in the phase diagram, as represented in  Fig. \ref{Fig:1}.  On cooling down, unconventional superconductivity with a $d$-wave symmetry occurs and seems to emerge out of the \gls{pg} state. Furthermore, deep inside the \gls{pg} state, an incipient  charge density wave (\gls{cdw}), with a $d$-wave structure factor,  competes  with superconductivity \cite{Keimer15,Proust19}, while strong antiferromagnetic (\gls{af}) spin fluctuations are still present in the materials. The incipient \gls{cdw} breaks the lattice translation (\gls{lt}) symmetry and gives rise to a new periodicity  associated with its modulation wave vector {\bf q}$_{c}\ne 0$. It coexists with a set of  intra-unit-cell (\gls{iuc})  electronic instabilities which preserve the \gls{lt} invariance ({\bf q}=0) (see {\it e.g.}  \cite{Fauque06,Fujita14,Fujita14b}), but break discrete Ising-like symmetries: the fourfold rotation (R) symmetry for an electronic nematic state  and additionally,   time reversal  (T) and parity (P) symmetries, as we shall discuss in details here. Remarkably, these different \gls{iuc} orders show up concomitantly with the \gls{pg} state. They are then necessarily tied up with 
it and should also be associated with a significant energy condensation. It should be stressed out that no additional noticeable change of the electronic entropy is reported below T$^*$, of course  apart from the one associated with superconductivity \cite{Loram01}. This is, for instance, the case for states like the \gls{cdw} occurring  within the \gls{pg} state.   A central question is therefore which one of these order parameters is the primary one ?  It should 
be also stressed that there is two known mechanisms to open gaps in metals, either it results from a finite {\bf q}$\ne$0 ordering breaking \gls{lt} or it is due to a particle-particle pairing as does superconductivity. Therefore, none of the \gls{iuc} broken symmetries alone can induce the needed fermionic gap characterizing the \gls{pg} state. 

To get a better understanding about these \gls{iuc} orders, it is useful to look at the basic atomic structure of superconducting cuprates. They  are layered perovskite materials,  made of the stacking of $\rm CuO_2$ layers, whose building block is a $\rm  CuO_2$ squared plaquette  (Fig. \ref{Fig:3}.a-c). In the insulating undoped state, there is only a single electron 
on the Cu site carrying a half-integer spin S=1/2. Upon hole doping, an 
extra hole on the O site can form a bound state with the electron on the Cu site, yielding the so-called Zhang-Rice singlet in the $\rm CuO_2$  plaquette\cite{Zhang88}. Owing to the strong on-site Coulomb repulsion 
which prevents a second hole to enter the plaquette, one is left with a set of three states for the  $\rm  CuO_2$ squared plaquette, useful to build a minimum model (such as the widely studied $t-J$ model) for cuprates, 
but which also eliminate the internal degrees of freedom within the $\rm CuO_2$  plaquette. In contrast, the \gls{iuc} electronic orders imply the emergence of non-Zhang-Rice singlet states\cite{Moskvin12}. For an electronic nematic state, an unbalanced electronic density at the O sites breaks the fourfold \gls{rotation}- symmetry of the $\rm CuO_2$ plaquette, as observed in scanning tunneling microscopy (\gls{stm})\cite{Fujita14,Fujita14b}.

Put altogether, the \gls{iuc} symmetry breakings point towards novel electronic states that have been proposed either to account for or to characterize
 the \gls{pg} state. That takes in particular the forms of loop currents (\glspl{lc}) circulating coherently between copper and oxygen orbitals \cite{Varma20}  as depicted in Fig. \ref{Fig:2}.a-b. 
 \gls{lc} states can take different shapes \cite{Varma97,Simon02,Varma06,Weber09,Weber14,Varma19}, where the electronic density remains uniform, but spontaneous charge currents appear between Cu and O sites within the $\rm CuO_2$ plaquette. The loop currents flow coherently among the three orbitals of copper and oxygen,  spontaneously breaking  \gls{time}-, \gls{parity}- and/or \gls{rotation}- symmetries. The search for  states of matter involving  charge currents  is a long story in the physics of high temperature superconductors and various states of matter were proposed. Unlike \gls{iuc} and \gls{lc} states (both being {\bf q}=0 electronic instabilities),  charge current density wave ({\bf q}$\neq$0 electronic instability) were also proposed \cite{Wang14}. For instance, within a single band Hubbard model, charge currents were predicted to flow around the CuO$_2$ square lattice, yielding the so called « $\pi$- flux phase » or D-density wave (DDW) \cite{Hsu91,Chakravarty01} which doubles the unit cell.  As the  $\pi$-flux phase breaks the \gls{lt} symmetry, additional superstructure  Bragg reflections are expected at 
several {\bf Q}-points in the reciprocal space equivalent to $(\pi,\pi)$ \cite{Hsu91,Chakravarty01}. Despite various attempts in different sets of 
samples, neutron diffraction measurements failed to prove the existence of magnetic peaks at the planar wave vector $(1/2,1/2)\equiv (\pi,\pi)$ in 
the 2D first Brillouin zone. As a generic property, all the types of charge current phases break \gls{time}-symmetry and closed current loops  generate a magnetic flux and orbital magnetic moments. That is the property primarily 
reported to provide evidence in favor of the \gls{lc} states in various quantum 
materials, and which people have tried to evidence by using dichroic effect using circularly polarized angle resolved photoemission (\gls{arpes}) \cite{Kaminski02} and next by looking for the \gls{lc} induced orbital magnetism using polarized neutron diffraction (\gls{pnd}) as we shall review here in {\bf section \ref{PND}}. To date, the \gls{lc} state has been reported in several oxide materials either with 2D structure, in the superconducting cuprates  \cite{Fauque06,Mook08,Baledent11,Mangin15,Mangin17,Li08,Li11,Tang18,Baledent10,deAlmeida12,Mangin14} and in the iridates  \cite{Jeong17}, or in the 1D spin ladder cuprate \cite{Bounoua20}, showing the universal nature of this exotic state. 

  In this article,  we next list  in {\bf section \ref{symmetry}} the various experimental data (other than \gls{pnd}) that support the emergence of such  \gls{lc} states built from the \gls{lc}-$\Theta_{II}$ pattern represented in Fig. 
\ref{Fig:2}.a. In particular, the macroscopic torque measurements in two cuprates families, namely  in bilayer YBa$_2$Cu$_3$O$_{6+x}$ and monolayer HgBa$_2$CuO$_{4+\delta}$,  \cite{Sato17,Murayama19}, although showing different results, match nicely the \gls{pnd} results in both materials \cite{Mangin17,Li08}. 

	The remainder of the manuscript focuses  in {\bf section \ref{modelPG}} on the interplay with the \gls{pg} state, which depends on the hierarchy among the various co-existing/competing observed phases.  \glspl{lc}  can be bound to the \gls{pg} physics in two ways. First,  the \gls{lc} phase is degenerate and several states corresponding to distinct \gls{lc} patterns are allowed \cite{Varma06}. This degeneracy can be the source of a proliferation of randomly distributed \gls{lc} domains. Assuming the existence of a super-cell made of a coherent juxtaposition of \gls{lc} domains\cite{Varma19} (as shown by Fig. \ref{Fig:2}.d-e),  the \gls{lt} invariance would be broken  and this modulated ({\bf q}$\rm  \neq $0)  \gls{lc}-based phase could trigger a gap opening on the Fermi surface. Following a fundamentally different approach,  the \gls{lc} order can also be understood in terms of a so-called "ancillary state" inseparable from the \gls{pg} state \cite{Agterberg15,Dai18,Scheurer18,Sarkar19}. The ancillary state is a secondary instability associated to a mother instability which itself effectively opens the \gls{pg}. For instance, it can correspond to pair-density-wave (\gls{pdw}) states \cite{Agterberg15,Dai18,Sarkar19}. Within these approaches, it was first pointed out by Agterberg {\it et al} \cite{Agterberg15} that the \gls{lc} phase 
necessarily happens owing to the symmetry and therefore is essential to capture the specific nature of the \gls{pg}. We conclude in {\bf section  \ref{conclude}} with some perspectives for future experimental developments. 

\section{\label{PND}Polarized neutron diffraction experiments}

Hunting for \glspl{lc}, we have performed \gls{pnd} experiments on several correlated materials over the last 15 years. We here discuss layered perovskite materials with a (nearly) tetragonal  structure within the $\bf ab$ basal plane corresponding to the (CuO$_2$ or IrO$_2$) layers  (Fig. \ref{Fig:3}c),  stacked along the $\bf c$-axis. The structural unit-cell contains one or more layers, depending on the materials, yielding different $c$ lattice parameter. Examples of the corresponding crystallographic structure are shown in Fig. \ref{Fig:3}a-b.  The \gls{pnd} technique is a momentum-selective probe for magnetism.  We commonly write, see {\it e.g.} ref. \cite{Fauque06}, the measured momentum as ${\bf {Q}}=(H,K,L)$ in reduced lattice units $(2\pi/a, 2\pi/b, 2\pi/c)$ with $a\sim b\sim$3.85 \AA. Various attempts to observe the different orbital magnetic moments associated with the charge currents  ($\pi$-flux or  D-density wave) phase have been already discussed in  \cite{Sidis07,Bourges11}.

\subsection{The \gls{iuc} ({\bf q}=0) magnetism}

Unlike the $\pi$-flux phase, the staggered orbital magnetism induced by the Varma's \glspl{lc}   \cite{Varma97,Simon02,Varma06} is associated with a set of pairs of charge currents looping clockwise and anticlockwise within each unit cell, Fig. \ref{Fig:2}.a  \cite{Sidis13}. Such a \gls{lc} state preserves the \gls{lt} symmetry, so that, in a diffraction experiment  its magnetic Bragg reflections are superimposed to the atomic Bragg reflections. This makes the experiments particularly difficult, as the magnetic contribution is weak, and polarized neutrons are required to discriminate between the atomic and magnetic scattering on a Bragg reflection.  Indeed,  any source of magnetism in the sample is able to flip the neutron spin, changing 
the spin polarization of the scattered neutron beam. To determine a possible orbital magnetic order, one needs to measure the variation of the neutron polarization versus temperature. A long range magnetic order would give rise to an upturn of the so-called  inverse flipping ratio, $\rm 1/FR$, which measures the ratio of magnetic to nuclear scattering when the neutron polarization is set parallel to the scattering wavevector \cite{Fauque06,Bourges11}. A typical result is reported  in Fig.  \ref{Fig:3}.e where $\rm 1/FR$  ratio is shown versus temperature in the bilayer cuprate YBa$_2$Cu$_3$O$_{6.6}$ \cite{Mangin17}. Two Bragg peaks are measured: at large  {\bf Q}, (2,0,0), where the magnetic signal should vanish, an expected smooth thermal variation is reported whereas an upturn is observed around T$\simeq$ 235K at low {\bf Q}=(1,0,0). This  corresponds to the occurrence of a spontaneous magnetic signal. Assessment of the technical difficulties and limitations related to the \gls{pnd} experiments have been recently re-emphasized in \cite{Bourges19}. 

\begin{figure}[tbp]
\includegraphics[width=14 cm]{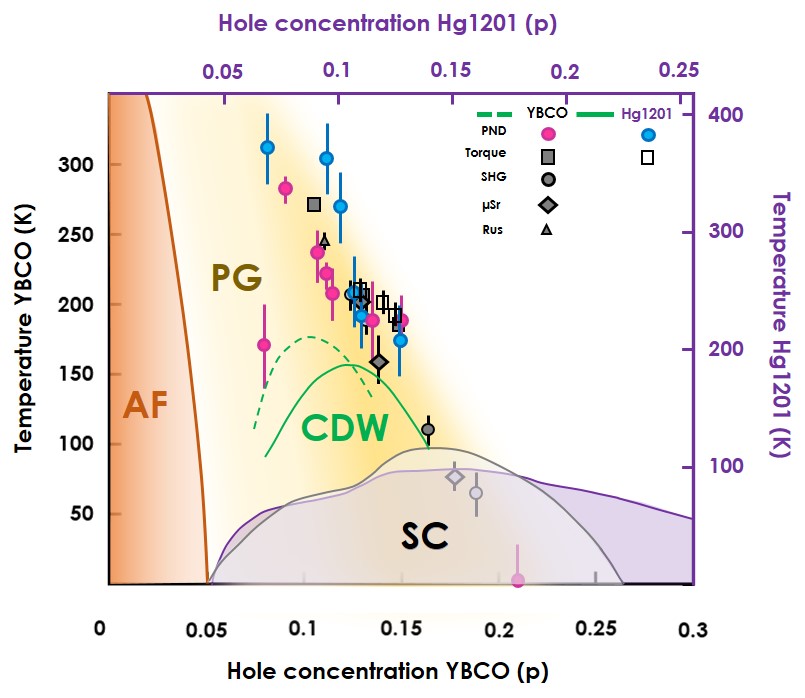}
\caption{ High-$\rm T_C$ superconducting cuprates phase diagram versus hole concentration for two cuprates families YBa$_2$Cu$_3$O$_{6+x}$ (YBCO) (bottom  and right scales) and HgBa$_2$CuO$_{4+\delta}$  (Hg1201)  (top  and left scales) showing the intra-unit cell (\gls{iuc}) magnetic order associated with the pseudogap state from polarized neutron diffraction (\gls{pnd}) \cite{Fauque06,Mook08,Baledent11,Mangin15,Mangin17,Li08,Li11,Tang18} and T$^*$, the pseudogap temperature, determined from torque \cite{Sato17,Murayama19}, second harmonic generation (\gls{shg}) \cite{Zhao16}, muon spectroscopy (\gls{musr}) \cite{Zhang18} and resonant ultrasound (Rus) \cite{Shekhter13} experiments. Representative lines of transition of antiferromagnetism (\gls{af}), superconductivity (SC), pseudogap (\gls{pg}) and charge density wave (\gls{cdw}) (full line for YBCO and dashed line for Hg1201) are also depicted. }
\label{Fig:1}
\end{figure}

Numerous \gls{pnd} measurements reveal the appearance of a {\bf q}=0 \gls{af} order at the Bragg reflections {\bf Q}=(1,0,L) \cite{Bourges11}, also referred to as an \gls{iuc} magnetism \cite{Sidis13}. This order displays a magnetic 
structure factor consistent with the orbital staggered magnetism produced 
by the \gls{lc}-$\Theta_{II}$ state within the CuO$_2$ planes. In the field of unconventional high temperature superconductors, we reported such observations in four distinct cuprate families : YBa$_2$Cu$_3$O$_{6+x}$ \cite{Fauque06,Mook08,Baledent11,Mangin15,Mangin17}, HgBa$_2$CuO$_{4+\delta}$ \cite{Li08,Li11,Tang18}, (La,Sr)$_2$CuO$_4$ \cite{Baledent10}, and Bi$_2$Sr$_2$CaCu$_2$O$_{8+\delta}$ \cite{deAlmeida12,Mangin14}. Fig. \ref{Fig:1} groups, in a generic phase diagram, the magnetic ordering temperatures  measured by \gls{pnd} in the monolayer  HgBa$_2$CuO$_{4+\delta}$ and bilayer YBa$_2$Cu$_3$O$_{6+x}$ compounds, as a function of the hole doping.

 Later, we also reported an \gls{iuc} magnetism in the large spin-orbit coupled Mott insulators, $\rm Sr_2(Ir,Rh)O_4$ monolayer iridate \cite{Jeong17},  suggesting a more generic character of such \gls{lc} orders in correlated electron oxides. It is worth noting that  both iridates (Ir, 5d) and cuprates (Cu, 3d) are both \gls{af}  insulators which can be doped (Mott or charge-transfer insulators, respectively, and S=1/2 on Cu versus J=1/2 on Ir). Likewise, they share a similar layered perovskite structure.  A noticeable difference with the 2D cuprates is that the \gls{lc} occurs as well  in the insulating compounds at half-filling, whereas in cuprates it is so far observed in the doped metallic and superconducting materials. It is worth stressing out that the mechanism that could be invoked to produce 
\gls{lc} state could differ in both oxides. 

Despite the large body of \gls{pnd} data supporting an \gls{iuc} magnetism in superconducting cuprates,  its existence was questioned by a \gls{pnd} attempt \cite{Croft17}, which failed to detect it in small YBa$_2$Cu$_3$O$_{6+x}$ single crystals (weighing less than 18 mg in contrast with our samples which weighted at least 2000 mg \cite{Bourges18}). Due to the limited flux of polarized neutron beams and the relatively weak orbital magnetic moments,  we demonstrated in \cite{Bourges18,Bourges19} that the experimental accuracy in  \cite{Croft17} was clearly insufficient to draw any conclusions from the data as a consequence of the one well-known drawback of neutron scattering, which is the systematic need of large enough single crystals. 

The \gls{pnd} results prove that the \gls{time}-symmetry is broken when entering the \gls{pg} phase of high-$\rm T_c$ superconducting  cuprates, consistent with the  \gls{lc}-$\Theta_{II}$ phase symmetry (Fig. \ref{Fig:2}.a) originally proposed by C.M. Varma \cite{Simon02,Varma06}. The deduced  spontaneous orbital magnetic moment is typically of the order of $\rm \sim 0.1 \mu_B$ per loop of the \gls{lc}-$\Theta_{II}$ phase \cite{Bourges11} and its magnitude decreases 
with increasing hole doping following the trend of the pseudogap  \cite{Fauque06,Mangin14}. In general, it should be stressed out that \gls{pnd} data can be interpreted by a large number of magnetic models as long as these models produce compensated magnetic moments within a given unit cell and respect the \gls{lt} symmetry.  This is the case for instance for the Dirac multipole ordering proposed in Ref.~ \cite{Lovesey15,Lovesey15b,Fechner16}, which we will  discuss below. To conclude, a model accounting for the \gls{pnd} data needs at least two opposite magnetic moments in each unit cell, to comply with the {\bf q}=0 \gls{af} structure.

\subsection{Magnetic moment direction}

\begin{figure}[tbp]
\includegraphics[width=12  cm]{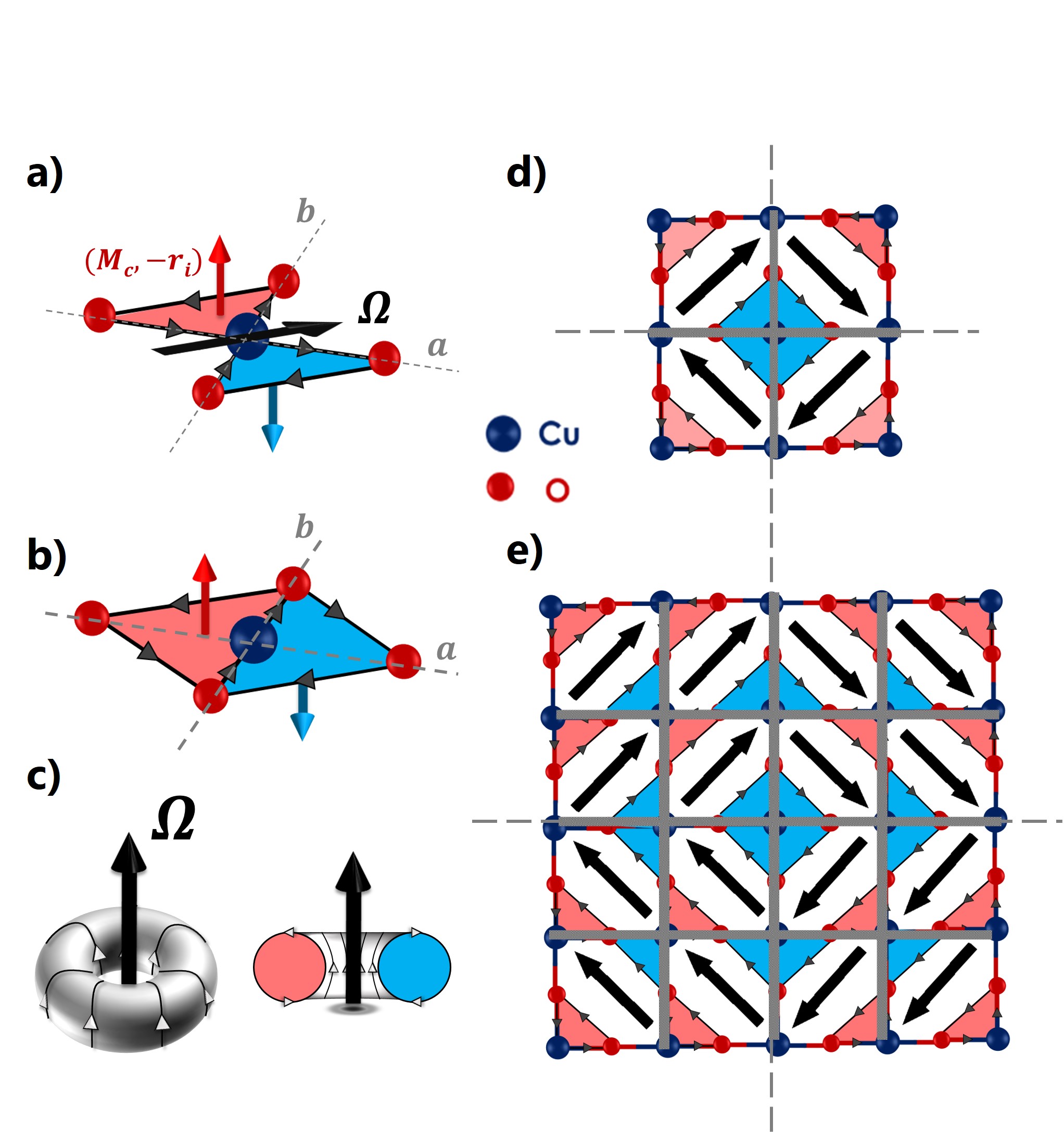}
\caption{  (a) Loop currents (\gls{lc}) in CuO$_2$ plane, so-called \gls{lc}-$\Theta_{II}$ 
phase, as proposed by C.M. Varma in superconducting cuprates \cite{Simon02,Varma06}, with the corresponding toroidal moment, $\Omega$, in black. (b) Model of loop currents proposed in Ref. \cite{Scheurer18}. 
 The \glspl{lc} now only run between oxygen sites, in contrast with the  \gls{lc}-$\Theta_{II}$ phase. (c) Anapole or polar toroidal moment ${\bf \Omega}$  formed by winding current around a solenoid. A 2D section in the torus reveals two loop currents turning clockwise (blue) and anti-clockwise (red). (d) $ (2P \times 2 P)$  \gls{lc}-$\Theta_{II}$ super-cell \cite{Varma19} formed by a $\pi/2$  rotation of the anapole moment (in black) between neighbouring unit-cells. The magnetic cell encompasses the 4 possible domains for a \gls{lc} pattern separated by domain walls (represented by the shaded lines). 
Here, the domains size as a function of the unit-cell is $P=1$ . (e)  $ 
(2P \times 2 P)$ magnetic supercell for  
$P=2$.}
\label{Fig:2}
\end{figure}

The direction of the magnetic moments is experimentally determined through the neutron polarization analysis \cite{Bourges11}, from which one can deduce the magnitude of the out-of-plane component ($\rm {\bf M}_{c}$) and the in-plane component ($\rm {\bf M}_{ab}$). In the  \gls{lc}-$\Theta_{II}$ phase, the orbital magnetic moments should be perpendicular to the $\rm CuO_2$ plane in which \glspl{lc} are confined (Fig. \ref{Fig:2}.a-b). However, as sketched in Fig. \ref{Fig:3}.d, the outcome of the \gls{pnd} analysis is a  $\Theta$$\sim 40 \pm 20 ^\circ$ 
tilt angle of the magnetic moment with respect to the {\bf c} axis. While 
systematically observed in all 2D cuprates \cite{Tang18}, $\Theta$ can vary noticeably between the cuprates, as well as a function of the measured 
momentum {\bf Q}-points. Interestingly, the moment tilt could also be temperature dependent \cite{Mangin15}. Furthermore, as  sketched in Fig. \ref{Fig:3}.d, the exact orientation of $\rm {\bf M}_{ab}$  in the $ab$ plane has not been determined experimentally yet. More generally, the observed tilt is not well accounted for by most of the  theoretical models proposed for the \gls{iuc} magnetism, as we shall see below. As a matter of fact, the observed tilt  is certainly a key and unexpected feature of the observed \gls{iuc} magnetism, putting stringent constraints on theoretical models.

\subsection{\gls{iuc} spatial correlations}

Before, going further in the discussion, it is worth emphasizing  some experimental limitations of the neutron scattering technique.  Generally speaking, the neutron scattering cross-section is convoluted with the instrumental resolution of the instrument in 4 dimensions ({\bf Q,} $\rm \omega)$ \cite{Hennion10}.  The determination of the correlation length and characteristic time scale are limited by the instrument resolution. In order to detect a weak magnetic response, one usually relaxes the resolution. 
 Likewise, a modulated structure, with a very long pitch, gives rise to satellite reflections very close to the Bragg spot. These satellite scatterings can merge with the Bragg scattering owing to the experimental resolution.  The canonical example is the  long period helimagnet MnSi \cite{Ishikawa1976}, for which the observation of the  magnetic satellites requires the use of a very good resolution for diffraction measurements or the 
use of small angle neutron scattering  techniques.  In the latter case, the measurement is  not performed  around a Bragg reflection at finite $\rm {\bf Q} ( \neq 0 )$ anymore, but around $\rm {\bf Q} \rightarrow 
0$.   For \gls{lc}-like magnetism, whose magnetic structure factor cancels at small angle,  small angle neutron scattering  is unfortunately unsuited.

In three of the cuprates families, namely  YBa$_2$Cu$_3$O$_{6+x}$, HgBa$_2$CuO$_{4+\delta}$ and Bi$_2$Sr$_2$CaCu$_2$O$_{8+\delta}$, the observed \gls{iuc} magnetism superimposes to the atomic Bragg peak \cite{Fauque06,Li08,Bourges11,deAlmeida12}, suggesting  that the \gls{iuc} order is long range and 3D 
as it is clearly established in underdoped YBa$_2$Cu$_3$O$_{6.6}$ \cite{Mook08}. Due the relaxed instrument resolution that was used, only an upper limit of the correlation length along the c axis $\xi_c \le $ 75 \AA\  could be given \cite{Mook08,Bourges11}. As in MnSi, a situation with magnetic satellites with a long magnetic pitch, $\lambda$, could happen. Various attempts 
to observe a magnetic response in the reciprocal space distant from the Bragg position were however  unsuccessful, giving a lower limit of a possible pitch of $\lambda \gtrsim 40$\AA. In an nearly optimally doped YBa$_2$Cu$_3$O$_{6.85}$ sample only, an \gls{iuc} magnetic signal could be detected broader than the in-plane momentum resolution \cite{Mangin15}. This clearly indicates short range correlations with finite in-plane correlation 
lengths, $\xi_{ab} \sim 20 a$ in that compound. 

A particularly interesting case is that of $\rm La_{2-x}Sr_xCuO_4$, a system which  is known to experience a segregation of doped holes into charge stripes  \cite{Julien03}. For Sr substitution close to x=1/12, we observed a  quasi-static  \gls{iuc} magnetic signal  that remained quasi-2D   with short range correlation lengths \cite{Baledent10}. This contrasts  with observations in other cuprates where a (quasi) long range \gls{iuc} magnetism develops at 3D. This short range magnetism in lightly doped $\rm La_{2-x}Sr_xCuO_4$ can be associated with very weak correlation lengths of \gls{lc}-$\Theta_{II}$ 1D ribbons ($\rm \sim 10$~\AA, i.e 2-3$a$). The  short range magnetism can be viewed as \gls{lc}-$\Theta_{II}$ objects confined on the charge stripes separated by hole poor \gls{af} domains of Cu spins, in relation with  the charge order in this material  \cite{Julien03,Bourges11}. \glspl{lc} can therefore interact with other electronic instabilities and persist in low dimensional systems (1D). 

Following this observation and theoretical predictions \cite{Chudzinski08,Scheurer18}, we recently studied the 2-leg ladder cuprate Sr$_{14-x}$Ca$_x$Cu$_{24}$O$_{41}$, which is an archetype of 1D spin liquids. This aperiodic system hosts a very rich phase diagram where, depending on the Ca content (which induces hole doping), it exhibits a strong spin liquid state ending into an intriguing ordered magnetic state at larger Ca content, passing through a \gls{cdw} phase. Using \gls{pnd}, we discovered the existence of 2D 
 short range magnetism in this material for two Ca contents \cite{Bounoua20}. This magnetism cannot be described by any type of Cu spins orders. Instead, our investigations establish the existence of \gls{lc} correlations within the 2-leg  ladders \cite{Bounoua20}.  The results from both studies suggest that \glspl{lc} could be confined within quasi-1D structures, such as  2-leg Cu ladders, imposed either by the structure of the material in Sr$_{14-x}$Ca$_x$Cu$_{24}$O$_{41}$ or by an electronic phase segregation  yielding the formation of 
(bond centered-) stripes in $\rm La_{2-x}Sr_xCuO_4$.

\subsection{$L$-dependence of the \gls{iuc} magnetic structure factor}

L dependence of the magnetic scattering associated with the \gls{iuc} order has 
been reported for wavevectors {\bf Q}=$(1,0,L)$ in two monolayer compounds  HgBa$_2$CuO$_{4+\delta}$\cite{Li08} and $\rm La_{2-x}Sr_xCuO_4$\cite{Baledent10}. In the former, the magnetic signal from \gls{iuc} order is only at integer $L$ values, while it remains 2D  (any $L$ values)  for the latter. Surprisingly, the scattered magnetic intensity displays the same fast 
decay as a function of $L$  \cite{deAlmeida12}. In principle, in bilayer compounds,  in-phase or out-of-phase coupling between layers would give rise to a magnetic structure factor modulated either by a term $4 \cos^2(\pi z L)$ or $4 \sin^2(\pi  z L)$, where $z$ (=d/c) stands for the reduced interlayer distance. Instead, the \gls{iuc} magnetic pattern in YBa$_2$Cu$_3$O$_{6+x}$  exhibits a crisscrossed stacking of the \gls{iuc} patterns \cite{Mangin17} (as illustrated in Fig. \ref{Fig:5}.b), so that the $L$-dependence associated with the bilayer vanishes in the magnetic structure factor when dealing with a twinned crystal. Accordingly, the $L$-dependence of the scattered magnetic intensity measured in bilayer systems like  twinned YBa$_2$Cu$_3$O$_{6+x}$ and Bi$_2$Sr$_2$CaCu$_2$O$_{8+\delta}$, matches that found for monolayer cuprates \cite{deAlmeida12}. Furthermore, the calibration of the magnetic intensity measured on Bragg reflections with similar |{\bf Q}| and in comparable samples ( i. e. similar  at T$^*$ and doping levels) indicates that the intensity is about only twice larger for system with 2 Cu per formula unit instead of 1 Cu /f.u.. This ratio is also consistent with the crisscrossed arrangement of \glspl{lc} within a bilayer. 

The fast decay of magnetic intensity along $L$ observed in all cuprates is actually much faster than the one controlled by |f({\bf Q})|$^2$ the squared magnetic form factor of Cu or O single ion \cite{deAlmeida12}.  This phenomenon could then be accounted for by \gls{iuc} magnetic moments being spread out within the CuO$_2$ layer.  In this picture, \glspl{lc} would not be simply confined within the CuO$_2$ layers. They would be more delocalized than a magnetic moment bound to a single ion, giving rise to a faster decay 
of the \gls{iuc} structure factor along the out-of-plane direction. This interpretation seems supported by the study carried out on twin-free samples: in this case, the scattered intensities along $(1,0,L)$ and $(0,1,L)$ differ and are weighted by $4 \cos^2(\pi z L)$ and $4 \sin^2(\pi z L)$, respectively  \cite{Mangin17}. However, the observed ratio between the magnetic \gls{iuc} intensities measured along a and b is not quantitatively reproduced with the value $z$=0.28. A larger value  of $z$ provides a better result, suggesting that the magnetic moments could be located outside the CuO$_2$ layer. Such a property has not been verified yet in all materials exhibiting \gls{iuc} order, such as iridates or 2-leg ladder cuprates.

\begin{figure}[tbp]
\includegraphics[width=12  cm]{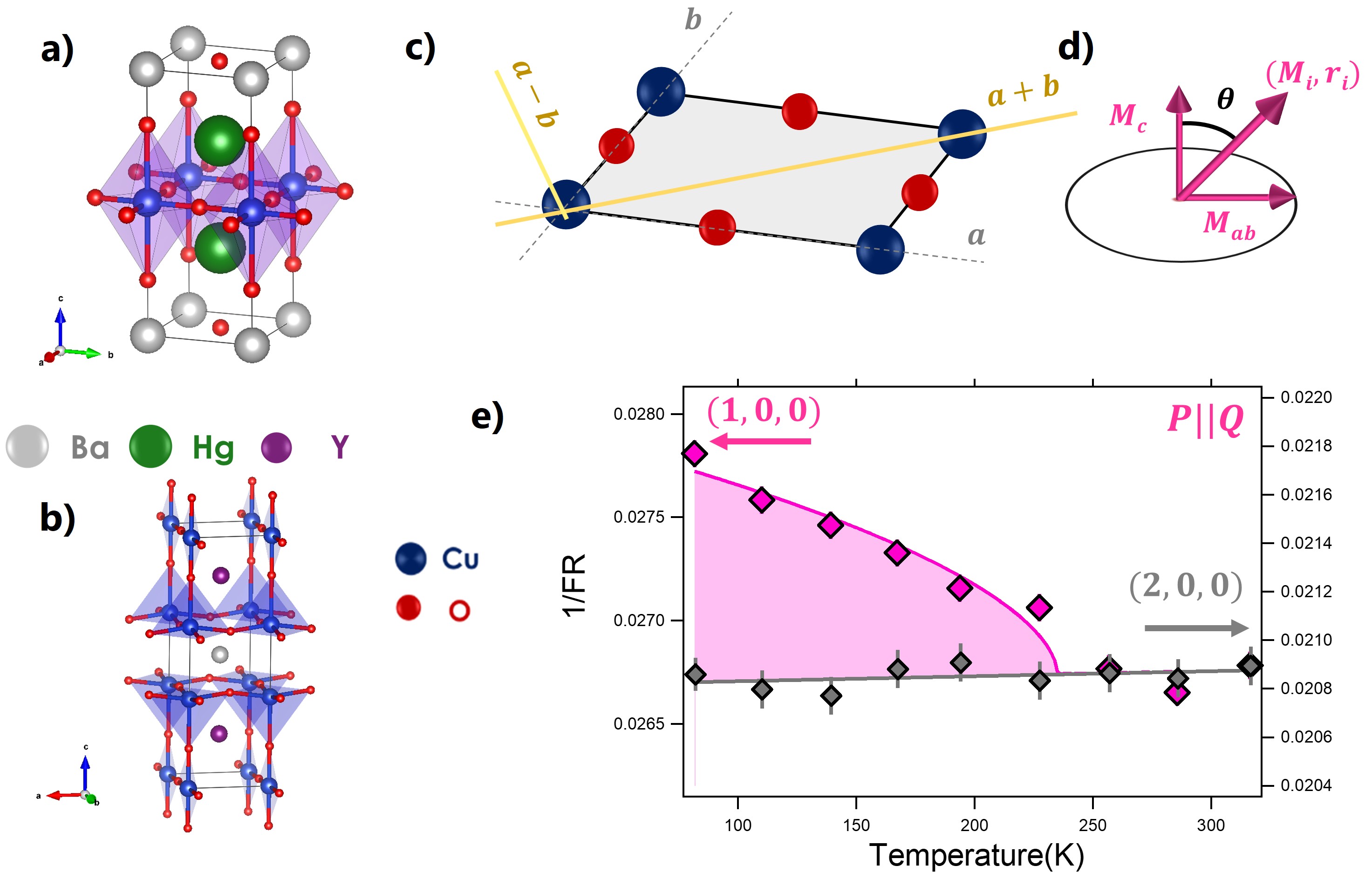}
\caption{Crystal structure of layered perovskites (a) HgBa$_2$CuO$_{4+\delta}$ and (b) YBa$_2$Cu$_3$O$_{6+x}$. (c) CuO$_2$ plane with copper (blue 
circles) and oxygen (red circles) atoms. (d) Tilt of the observed magnetic moment. The magnetic moment is tilted by an angle $\theta$ from the c-axis direction, perpendicular to the CuO$_2$ plane. (e) Temperature dependencies of the inverse of the flipping ratio $\rm 1/FR$ (see text) measured at $\rm {\bf Q}$ = (1,0,0) (purple) and at $\rm {\bf Q}$ = (2,0,0) (gray) from  \cite{Mangin17}.}
\label{Fig:3}
\end{figure}

\section{ \label{symmetry} Symmetry considerations}

In both cuprates and iridates, \gls{pnd} shows the existence of an \gls{iuc} magnetism, whose characteristic properties are consistent with  the  \gls{lc}-$\Theta_{II}$  phase pattern breaking  discrete \gls{time}-, \gls{parity}- and  fourfold \gls{rotation}- symmetries. In addition to \gls{pnd}, complementary evidences of such  broken symmetries were reported by other techniques, ranging from  dichroism in circularly polarized \gls{arpes} spectroscopy \cite{Kaminski02}, Kerr effect \cite{Xia08},  Nernst effect \cite{Daou10}, optical birefringence measurements \cite{Lubashevsky14}, second harmonic generation (\gls{shg}) \cite{Zhao16,Zhao17},  torque \cite{Sato17,Murayama19,Murayama20} and recently photo-galvanic effects \cite{Lim20},  respectively. The Kerr effect was reported in cuprates and necessarily implies a global \gls{time}-broken symmetry although this happens in $\rm YBa_2Cu_3O_{6+x}$  \cite{Xia08} at a slightly lower temperature than the other experimental probes \cite{Fauque06,Daou10,Lubashevsky14,Zhao16,Sato17}. Each technique brings complementary information  to draw the portrait of the \gls{iuc} order and to guide theoretical models.

\subsection{Anapoles and quadrupoles}

The  \gls{lc}-$\Theta_{II}$ state is the ground reference of the observed \gls{iuc} magnetic order. It is characterized by \gls{time}- and \gls{parity}-  symmetry breaking (although their product is preserved) \cite{Simon02,Varma06}.  This view was reinforced in cuprates \cite{Zhao17} and put forward in iridates \cite{Zhao16} by the observation of \gls{parity}-symmetry breaking in \gls{shg} experiments at the same temperature for a given doping. The  \gls{lc}-$\Theta_{II}$ magnetic pattern is described by  two staggered orbital magnetic moments $\rm {\bf M_i}$ 
 located on a diagonal of the $\rm CuO_2$  plaquette  and  equidistant from the Cu site. Within the basal lattice and  for the state described in Fig. \ref{Fig:2}.a,  the magnetic moments $\rm {\bf M}_i= \pm (0,0,\pm M_c)$ are then located at  $\rm {\bf  r}_i= \pm (-r_0, r_0,0)$. The  \gls{lc}-$\Theta_{II}$ exhibits, in principle, magneto-electric effects that can be described by a uniform arrangement of  anapoles (also named : toroidal 
 moments) \cite{Shekhter09}. The anapole $\bf {\Omega}$ corresponds to a polar vector shown in Fig. \ref{Fig:2}.a constructed from the \gls{lc} as,

\begin{equation}
{\bf \Omega}=\sum_i  \bf{r_i} \wedge \bf{M_i}
\end{equation}

In general,  an anapole is described using a current flowing around a solenoid bent to form a  torus  (Fig. \ref{Fig:2}.c). The winding current around the solenoid generates a circular magnetic field inside the torus which  gives rise to a pure anapole along the axis of the torus, according to the definition of ${\bf \Omega}$. In 2D, the 3D torus reduces to 2  \glspl{lc} turning clockwise and anticlockwise. Actually, this staggered orbital magnetism supports not only an anapole, but also a magnetic quadrupole tensor \cite{DiMatteo12}. This is not surprising as the anapole appears at the same order of the magnetic multipole expansion as the magnetic quadrupole moment, as discussed in the context of multiferroics \cite{Spaldin08,DiMatteo12}.  More precisely, for the \gls{lc}-$\Theta_{II}$ state, $ \rm {\bf \Omega}= 2 r_0 M_c ({\bf a+b})$  and the  non-zero components of the magnetic quadrupole  $ \rm Q_{ac}= Q_{bc}=   r_0 M_c$ \cite{DiMatteo12}. 


\gls{lc}-free models, involving Dirac multipoles, were actually developed  to account for the \gls{iuc} magnetic scattering reported by \gls{pnd}  \cite{Lovesey15,Lovesey15b,Fechner16}. The $\rm CuO_2$ plaquette is not occupied anymore by an anapole bound to \glspl{lc} flowing through Cu and O, but by magnetic quadrupoles localized on the Cu site. Symmetry arguments are used to select the allowed magnetic quadrupoles, which can further ensure the \gls{parity}- and \gls{time}- symmetries breaking \cite{Lovesey15,Lovesey15b,Fechner16}.  
Unlike electronic multiferroics where the charge and magnetic degrees of freedom form two distinct order parameters,  these theoretical models predict the existence of new magneto-electric objects, in which the charge and electronic degrees of freedom are intimately coupled to form a single exotic order parameter, for instance due to electron-phonon coupling  \cite{Fechner16}.

Similarly to the magnetic distribution of the \gls{lc}-$\Theta_{II}$ state, a magnetic distribution with only in-plane magnetic moments can be considered as well \cite{DiMatteo12}: this corresponds to a pure magnetic quadrupole which is generated by a set of magnetic dipoles.  Keeping the same notation as for the \gls{lc}-$\Theta_{II}$ state, one could consider two magnetic moments $\rm {\bf M}'_i= \pm ( -M_{ab}, M_{ab},0)$,  again located at  $\rm {\bf  r}_i=\pm (-r_0,r_0,0)$. The  non-zero components of the magnetic quadrupole are  $ \rm Q_{aa}=\frac{1}{3}Q_{ab}= -\frac{1}{2}Q_{cc}= \frac{2}{3}  r_0 M_{ab}$ \cite{DiMatteo12} and differ from the ones 
of the quadrupole above. These planar magnetic moments do not generate an 
anapole. The existence of such a magnetic  quadrupole could then explain the observed in-plane component ($\rm {\bf M}_{ab}$) of the \gls{iuc} magnetism 
reported by \gls{pnd}, whereas the original \gls{lc}-$\Theta_{II}$ state can only explain the measured out-of-plane component ($\rm {\bf M}_c$). As one observes experimentally a tilted \gls{iuc} moment, none of these magnetic distributions alone can describe the \gls{pnd} results. However, a combination of these two pictures could account for the tilted \gls{iuc} moment as we shall see below (section \ref{pointgroup}).

\subsection{Degeneracy of \gls{lc}-$\Theta_{II}$ state}

It is worth emphasizing that the \gls{lc}-$\Theta_{II}$ phase is characterized by four degenerate quantum states.  The four \gls{lc} patterns  are deduced from each other by a 90$^{\circ}$ rotation with the anapole pointing along all diagonals of the basal plane \cite{Aji10,Ye12}. They produce four types of  magnetic domains that will be summed up in bulk \gls{pnd} experiments (Fig. \ref{Fig:3}.e). If one considers a classical \gls{lc} model, each of the \gls{lc} domains gives a magnetic scattering in \gls{pnd} corresponding to magnetic moments perpendicular to the $\rm CuO_2$ layer.  In contrast, if a quantum superposition of the four \gls{lc} states is allowed, then an effective planar magnetic response can appear as a result of quantum interference in the neutron scattering cross-section \cite{Ye12}. Therefore, the existence of the in-plane magnetic response could highlight the quantum effects in the \gls{lc} state. The observation of a temperature dependent tilt angle in nearly optimally doped YBa$_2$Cu$_3$O$_{6.85}$ \cite{Mangin15} could suggest a crossover from classical to quantum \gls{lc} states, when thermal fluctuations are overcome by quantum ones. Within that scenario, quantum effects would then occur at relatively high temperature, around 200-250 K. This might be surprising unless the \gls{pg} itself could impose the quantum superposition of the \gls{lc} states which in turn produce the observed tilt \cite{Ye12}.

\begin{figure}[tbp]
\includegraphics[width=12 cm]{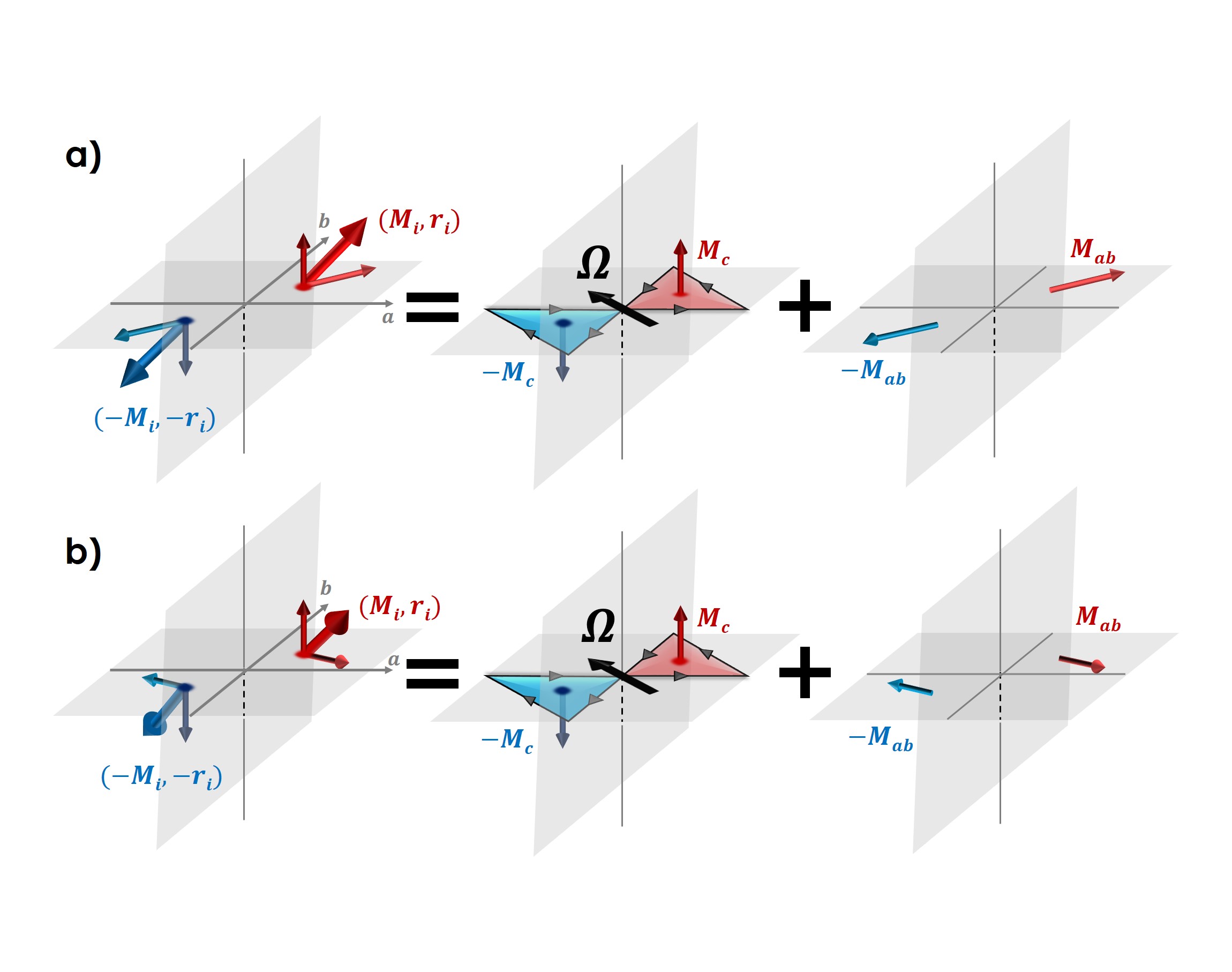}
\caption{Representation of two possible magnetic moment arrangements which are compatible with the \gls{pnd} results. They correspond to a superposition 
of out-of-plane moment as the \gls{lc}-$\Theta_{II}$ phase \cite{Simon02,Varma06} with in-plane moments either a)  along the diagonal joining up the moments, or b)   along the perpendicular diagonal. This yields to different symmetries with a)  $2/m'$ point group and b) $2'/m$ point group. Note that, as in Fig. \ref{Fig:2}.a, only one of the four possible \gls{lc} states is shown. 
 }
\label{Fig:4}
\end{figure}

\subsection{\gls{iuc} point group symmetry \label{pointgroup}}

Considering the simple crystal structure of a tetragonal  monolayer system, such as HgBa$_2$CuO$_{4+\delta}$  (see Fig. \ref{Fig:3}.a), the \gls{lc}-$\Theta_{II}$ state belongs to the magnetic point group $m'mm$ \cite{Varma06, Orenstein11} ($m'$ is a mirror plane followed by a time reversal operation).  The three mirror planes are respectively: the ({\bf a+b},{\bf a-b}), ({\bf a+b},{\bf c}) and ({\bf a-b},{\bf c}) planes.  Combining the above discussed magnetic distributions, one can build the magnetic pattern reported in  Fig.~\ref{Fig:4}.a  with  two magnetic moments $\rm {\bf M}_i= 
\pm (M_{ab}, M_{ab},M_c)$.  This magnetic arrangement corresponds to a rotation of the \gls{lc}-$\Theta_{II}$  magnetic pattern  around the  {\bf (a-b)} 
 diagonal axis, yielding a monoclinic magnetic point group $2/m'$ (as the 
rotation  preserves the $m'$ mirror plane)  \cite{Orenstein11}. The magnetic pattern shown in Fig.~\ref{Fig:4}.a  differs from the one shown in Fig.~\ref{Fig:4}.b where the planar magnetic component is aligned along the 
 {\bf (a-b)} axis (the moments are tipped out of the plane of the $M_c$  components).  The latter magnetic pattern has a magnetic point group $2'/m$  \cite{Orenstein11}.  Note that the planar magnetic components  of Fig.~\ref{Fig:4}.b cannot be associated with a pure magnetic quadrupole only as in Fig.~\ref{Fig:4}.a. Indeed, they give a contribution to an anapole  ($\rm -4 r_0 M_{ab} {\bf  c} $), in addition to  the magnetic quadrupole ($ \rm Q_{aa}=-Q_{bb}= 2  r_0 M_{ab}$). 
While being distinct, the magnetic patterns reported in Fig.~\ref{Fig:4} produce hardly distinguishable neutron scattering cross-section and fully 
account for the observed \gls{iuc} magnetism.  It should be stressed out that the $2'/m$ point group was deduced from \gls{shg} results in YBa$_2$Cu$_3$O$_{6+x}$ \cite{Zhao17} based on i) the observed $2/m$ symmetry of the atomic structure at 300K  and ii)  the broken global inversion related to the absence of C$_2$ axis. 

In order to account for the existence of an in-plane magnetic component, one could also relax the constraint that \glspl{lc} have to be confined within the $\rm CuO_2$ plaquette. In hole doped cuprates, the Cu site is located at the center of an O octahedron in monolayer compounds, which splits in O pyramids in bilayer ones.  The \glspl{lc} could then be delocalized on the faces of the CuO$_6$ octahedra  or on the CuO$_5$ pyramids, yielding a natural tilt of their orbital moments. However, the \gls{pnd} experiments exclude the \gls{lc} states occurring on the faces of the CuO$_6$ octahedra  \cite{Weber09,Orenstein11,Yakovenko15} or on the CuO$_5$ pyramids \cite{Lederer12} because the basal plane cannot also be a mirror plane for the in-plane component $\rm {\bf M}_{ab}$.  Only the \gls{lc} state  decorating the CuO$_6$  octahedra with only two opposite moments, as shown in Fig. \ref{Fig:5}.a and 
represented in \cite{Agterberg15,Yakovenko15}, is  compatible so far with 
the \gls{pnd} results in HgBa$_2$CuO$_{4+\delta}$ \cite{Li08,Li11,Tang18} and  belongs to  the $2/m'$ point group. It should be stressed that no \gls{lc} models on the CuO$_5$ pyramids have been found to be consistent with the \gls{pnd} data in bilayer materials. 

Beyond the \gls{lc} approach, a (Cu-)quadrupole modelling of the \gls{pnd} data \cite{Lovesey15,Lovesey15b,Fechner16}  is also proposed with  a magnetic structure belonging to $ m'm'm'$ point group. It is not consistent actually  with the data, as the basal plane cannot have a $m'$ symmetry plane for the $\rm {\bf M}_c$ component.  This is due to the measurement of the magnetic structure factor at the {\bf Q}-position for $L=0$ where a larger tilt is typically deduced \cite{Mook08,Tang18}.  Other possible quadrupole 
arrangements \cite{Fechner16} are then designed  to give a magnetic structure that belongs to the $2/m'$ point group.

In the  bilayer YBa$_2$Cu$_3$O$_{6+x}$ system, the \gls{pnd} results in a detwinned sample \cite{Mangin17} (see {\it e.g.} Fig. \ref{Fig:5}.b) are important as they rule out models \cite{Lovesey15,Fechner16,Lederer12}, which assume either an in-phase or out-of-phase coupling of \gls{iuc} order within a given bilayer. This dismisses models based on a magnetic nematic order, involving spin or orbital moments located at the O sites \cite{Fauque06,Moskvin12} as well. 

Linear and circular photo-galvanic effects recently observed at the \gls{pg} temperature \cite{Lim20} in both  monolayer and bilayer Bi-based cuprates suggest a $mm2$ point group symmetry, although monoclinic point group as $2/m'$ or $2'/m$ are also possible. A very interesting outcome of these experiments is the fact that the phase coherence of the \gls{iuc} order should be 
as large as the size of the illuminated beam ($\rm \sim 10 \mu m$).

\subsection{Rotational symmetry breaking}

The different \gls{iuc} phases proposed break \gls{rotation}-symmetry. This can be tested in 
various experiments as, for instance, via the loss of both C$_4$ rotation 
and mirror symmetry in the electronic structure of the  CuO$_2$ plane \cite{Lubashevsky14}. Indeed, the  \gls{lc}-$\Theta_{II}$ state imposes a preferential direction of the anapole along one of the diagonals of the CuO$_2$ plane.  On the opposite,  staggered \glspl{lc} flowing on O sites only, as proposed in Ref.~ \cite{Chatterjee17,Scheurer18},  give a pattern rotated at  45$^\circ$ with respect to  \gls{lc}-$\Theta_{II}$ one and are characterized by an anapole along one of the CuO bonds (Fig. \ref{Fig:2}.b). However, owing to the formation of degenerate domains and assuming equal populations of the domains, \gls{pnd} cannot distinguish between these different magnetic models, see {\it e.g.} our results in  2-leg ladder cuprates  \cite{Bounoua20}.  Other techniques can provide information on the \gls{rotation}-symmetry breaking. The broken fourfold  \gls{rotation}-symmetry is very often associated with the existence of electronic nematicity, which was reported using magnetic torque measurements in both cuprates  \cite{Sato17,Murayama19} and iridates \cite{Murayama20}. In the monolayer   HgBa$_2$CuO$_{4+\delta}$ compound, the torque magnetometry reports an anisotropy, $\chi_{ab}\ne 0$, along 
the diagonals of the CuO$_2$ plane \cite{Murayama19}, consistent with a distortion induced by the original \gls{lc}-$\Theta_{II}$ state \cite{Shekhter09}. In monolayer iridates, torque experiments also show an electronic nematicity  \cite{Murayama20} with a nematic director, which  is  rotated by  
45$^\circ$ from that of monolayer cuprate HgBa$_2$CuO$_{4+\delta}$. This implies the different  \gls{lc}  patterns between the two systems, suggesting  orbital-current configurations also turned by  45$^\circ$  (Fig. \ref{Fig:2}.b)\cite{Chatterjee17,Scheurer18}. 

\begin{figure}[tbp]
\includegraphics[width=12 cm]{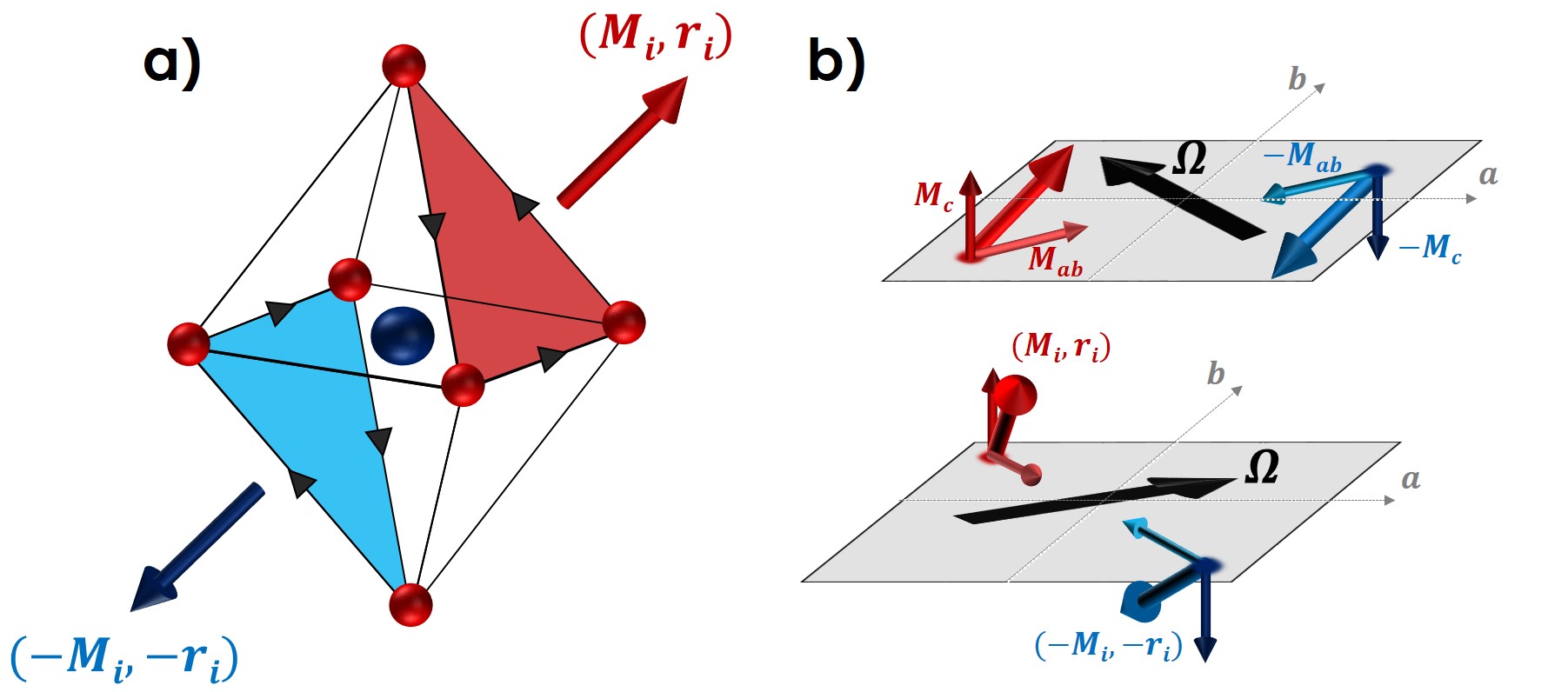}
\caption{(a) Loop currents decorating a CuO$_6$  octahedra with two circulating loops \cite{Agterberg15,Yakovenko15} which can describe the \gls{pnd} results in HgBa$_2$CuO$_{4+\delta}$ \cite{Li08,Li11,Tang18}. (b) Model of magnetic moments for a bilayer cuprate like YBa$_2$Cu$_3$O$_{6+x}$  built from the magnetic moment arrangement of a single layer shown in \ref{Fig:4}.a and compatible with the observed neutron structure factor \cite{Mangin17}.}
\label{Fig:5}
\end{figure}

In the bilayer YBa$_2$Cu$_3$O$_{6+x}$ compound, there is a weak orthorhombicity ($a\neq b $) and an electronic nematicity 
was reported through Nernst effect measurements \cite{Daou10}, showing the spontaneous appearance of a strong $ab$ anisotropy. In addition, the torque magnetometry shows an anisotropy along the principal directions, with $\chi_{aa} > \chi_{bb}$\cite{Sato17} {\it i.e.} turned by 45$^\circ$ from the single layer compound.  Actually, the surprising difference in torque measurement results in monolayer HgBa$_2$CuO$_{4+\delta}$  and bilayer YBa$_2$Cu$_3$O$_{6+x}$  compounds can be solved by a close examination of the way   \gls{lc}-$\Theta_{II}$ patterns are stacked within a bilayer (along the $\rm {\bf c}$-axis) as shown by previous \gls{pnd} results  \cite{Mangin17}. Indeed, using a twin-free sample of the orthorhombic YBa$_2$Cu$_3$O$_{6.6}$, our \gls{pnd} results show a distinct $ab$ anisotropy of the \gls{iuc} magnetic structure factor \cite{Mangin17}, as a result  of \gls{rotation}-symmetry breaking. 
This highlights that the \gls{iuc} order in this material breaks the mirror symmetry of the CuO$_2$  bilayer  (Fig. \ref{Fig:5}.b). For a single $\rm CuO_2$ plaquette, 4  \gls{lc} patterns are allowed and, then,  $4 \times 4$  \gls{lc} patterns  for a bilayer. Only 4 of them produce the magnetic structure factor  matching the observed $ab$ anisotropy found in \gls{pnd} measurements. They correspond to a crisscrossed  arrangement of the \gls{lc} patterns in the bilayer, yielding a resulting anapole, $\rm {\bf \Omega}_1 + {\bf \Omega}_2$, parallel to {\bf b} axis \cite{Mangin17}  as shown in Fig. \ref{Fig:5}.b. Note that each of the 4 selected \gls{lc} patterns could be fully identified 
when  using  both their anapole and their chiral parameter,  $\rm {\bf c } ({\bf \Omega}_1  \wedge   {\bf \Omega}_2)$.

\begin{figure}[tbp]
\includegraphics[width=12 cm]{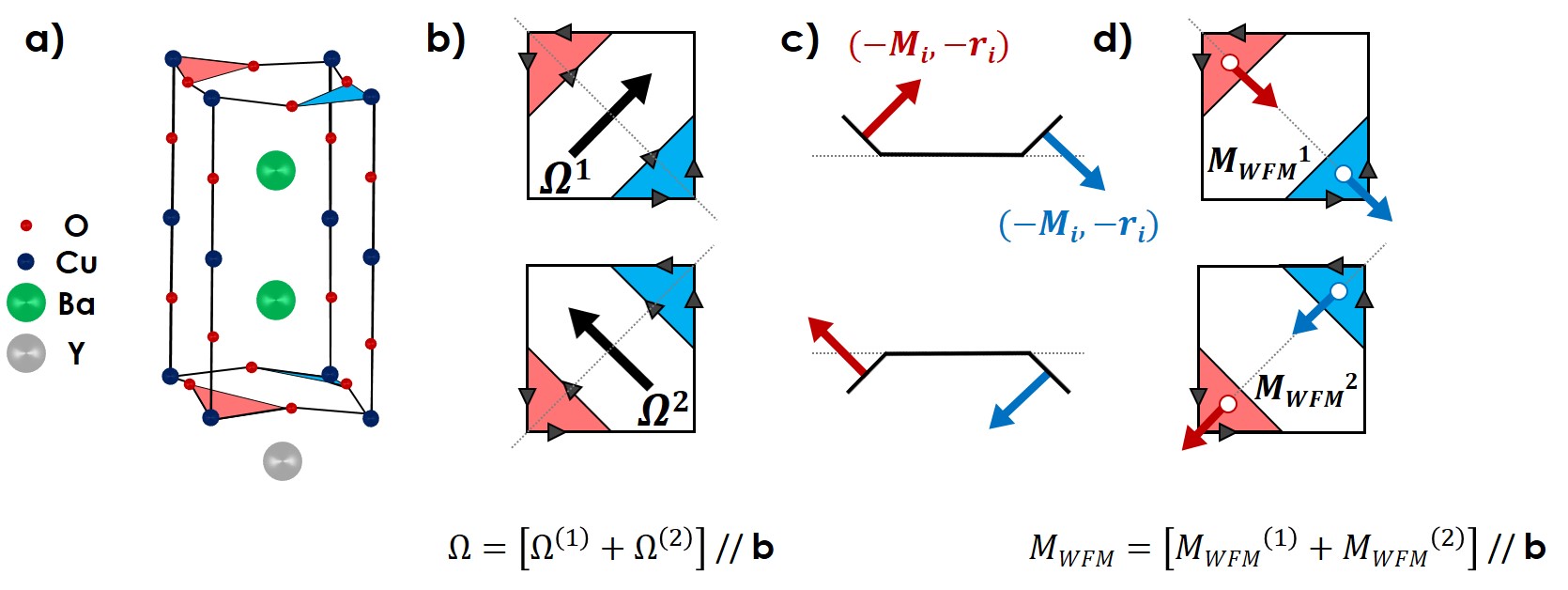}
\caption{(a) Crystal structure of YBa$_2$Cu$_3$O$_{6+x}$, showing the corrugation or dimpling of the CuO$_2$ planes. (b)  Top view of the loop currents arrangements of the \gls{lc}-$\Theta_{II}$-type in each  CuO$_2$  layer (from \cite{Mangin17}) where the toroidal moment is shown in black. The summed toroidal moment for a bilayer is pointing along the $\bf b$ direction.  (c) Cut of the unit cell along the diagonal showing the orbital moments perpendicular to the O-Cu-O triangle.  The dimpling of the O-Cu-O triangle gives rise to an horizontal magnetic component along the diagonal direction.   (d)  Top view of the summed magnetic moment, also pointing along $\bf b$.}
\label{Fig:6}
\end{figure}

 From this result and using the corrugation of the $\rm CuO_2$ plane (usually called dimpling in the literature), corresponding to the fact that Cu and O atoms lay in slightly different  (but parallel) planes, one can construct a picture to explain the origin of the torque anisotropy  (Fig. \ref{Fig:6}a-d). Assuming that \glspl{lc} run exactly  within the Cu-O triangle, 
the dimpling of the CuO$_2$ layer induces  a tilt of the out-of-plane moment, yielding a weak ferromagnetic in-plane component ${\bf M}$, pointing perpendicular to the anapole $\rm {\bf \Omega}_i$  in each layer.  
As the total bilayer anapole is found to be along the {\bf b} direction, the remaining  weak ferromagnetic  component for the bilayer is found also along   {\bf b} 
 (Fig. \ref{Fig:6}.d).
Accordingly, the magnetic torque $\tau$, defined as  $\tau \propto {\bf M} \wedge  {\bf H}$ for an applied magnetic field ${\bf H}$ along ${\bf c}$, is then expected to give an anisotropy along the {\bf a} direction as it is reported in \cite{Sato17}. This simple picture gives  a nice agreement between the microscopic picture deduced from \gls{pnd} experiment and the macroscopic torque measurements. It should be stressed  that the  weak ferromagnetic  component would be averaged owing to the existence of the four domains 
of the  \gls{lc}-$\Theta_{II}$ state \cite{Aji10,Ye12}.

Electronic nematicity was reported in the bilayer Bi$_2$Sr$_2$CaCu$_2$O$_{8+\delta}$  family as well, using  either \gls{stm} images of the \gls{iuc} states \cite{Fujita14,Fujita14b,Lawler10,Mukhopadhyay19} or electronic Raman scattering in the $B_{1g}$ symmetry \cite{Auvray19}. As in the bilayer YBa$_2$Cu$_3$O$_{6+x}$ material, a nematic director is found along the Cu-O bonds. Therefore, the nematic state in Bi$_2$Sr$_2$CaCu$_2$O$_{8+\delta}$ can be understood the same way assuming a similar crisscrossed arrangement of \glspl{lc} within the CuO$_2$ bilayer, {\it i.e.} with a total anapole moment 
in the bilayer pointing along the Cu-O bonds.  Alternatively,  the \gls{lc}-$\Theta_{II}$ state is a $\rm {\bf q}=0$ electronic state that could co-exist with other orthogonal $\rm {\bf q}=0$ electronic states, such as an 
 electronic nematic state.  

To conclude, the proposed  \gls{lc}-$\Theta_{II}$ represents the most adequate starting point to describe the \gls{iuc} magnetism observed in \gls{pnd} experiments. 
  It takes into account the various constraints imposed by symmetry even though additional features  need to be considered. 


\section{ \label{modelPG} Loop currents for Cuprates}

\subsection{Pseudogap line in cuprates: a phase transition}

\begin{figure}[tbp]
\includegraphics[width=12 cm]{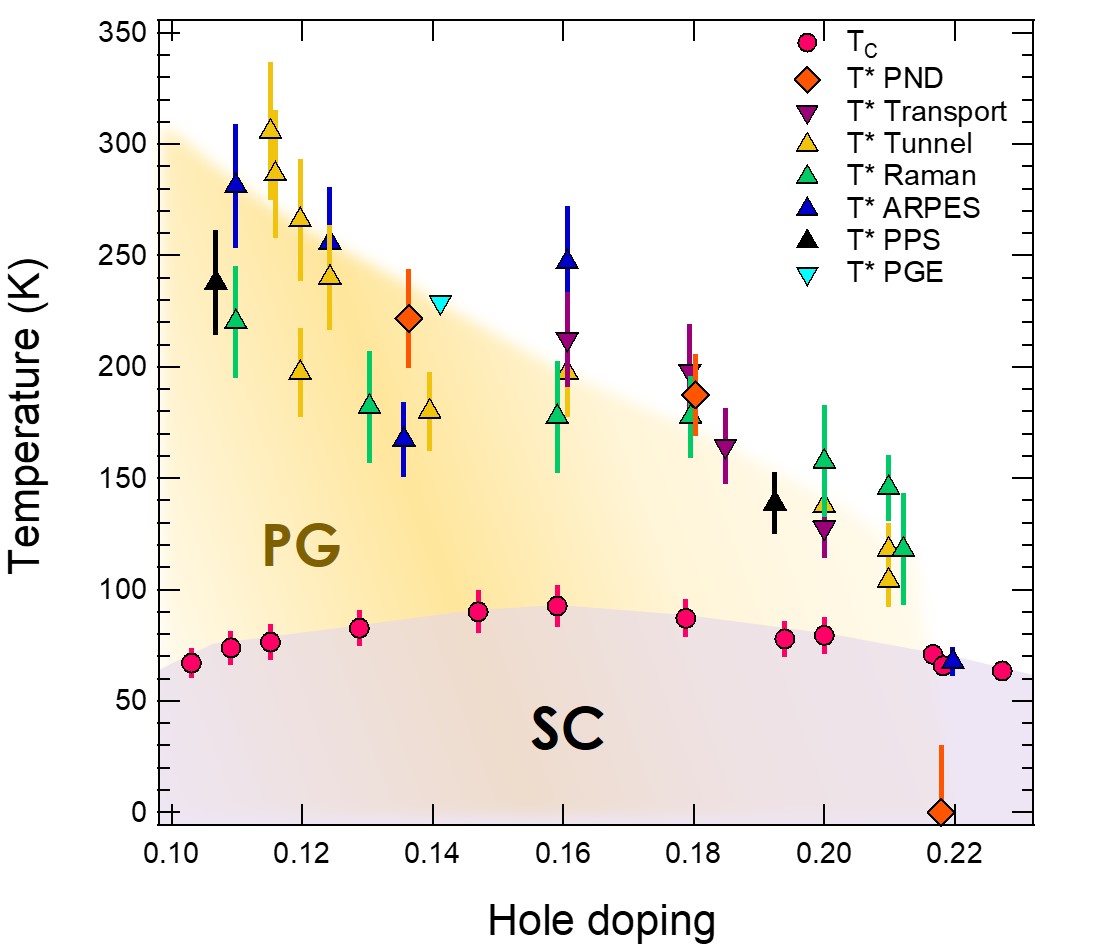}
\caption{ High-$\rm T_C$ superconducting cuprates phase diagram versus hole concentration for the cuprate family  Bi$_2$Sr$_2$CaCu$_2$O$_{8+\delta}$ showing the intra-unit cell (\gls{iuc}) magnetic order associated with the pseudogap state from polarized neutron diffraction (\gls{pnd}) \cite{deAlmeida12,Mangin14}, transport \cite{Watanabe97},  tunnel\cite{Dipasupil02,Ozyuzer02},  electronic Raman scattering \cite{Loret20},  \gls{arpes}\cite{Vishik12}, 
pump probe spectroscopy  (PPS) \cite{Toda04}  and photo-galvanic (PGE)  \cite{Lim20} experiments. Representative area for superconductivity (SC) and pseudogap (\gls{pg})  are also depicted. }
\label{Fig:7}
\end{figure}

Describing the generic experimental phase diagram shown in Fig. \ref{Fig:1}, one realizes  that the \gls{iuc} order occurs at T$^*$, the \gls{pg} onset temperature. 
The \gls{pg} energy scale impacts a large number of experimental spectroscopic probes in cuprates,  for instance electronic Raman scattering \cite{Sacuto13}. 
For a long time, the \gls{pg} physics in 2D cuprates has been considered as a crossover phenomenon observed in various physical properties, occurring over a certain range of temperatures. No sharp anomaly was seen in the specific heat data, from which it was concluded that T$^*$, where the \gls{pg} opens 
in transport measurements, cannot correspond to a thermodynamic phase transition \cite{Loram01}. In the recent years, as pointed out above, various symmetry breakings were identified over a much thinner range of temperatures. Independent measurements of an order parameter breaking \gls{time}-, \gls{parity}- and \gls{rotation}-symmetries agree on the location of a phase transition seen in  experiments probing the thermodynamic properties. This is clearly established in three high-$\rm T_c$ superconducting cuprate families (the monolayer  HgBa$_2$CuO$_{4+\delta}$, the bilayers  YBa$_2$Cu$_3$O$_{6+x}$  and Bi$_2$Sr$_2$CaCu$_2$O$_{8+\delta}$). Fig. \ref{Fig:7} shows exclusively experimental results in  Bi$_2$Sr$_2$CaCu$_2$O$_{8+\delta}$. That system is important to adress the issue of the \gls{pg}  because surface spectroscopic techniques can be performed as that material cleaves easily. Numerous \gls{arpes} \cite{Vishik12} and tunneling spectroscopy of junctions \cite{Dipasupil02,Ozyuzer02}  studies have been performed to establish the depletion of the electronic  states in that particular cuprate family. These estimates agree with T$^*$ deduced from resistivity measurements \cite{Watanabe97} as well with those obtained from various spectroscopies  \cite{Toda04,Loret20}. The \gls{iuc} magnetism  observed using \gls{pnd} was among the first reported order parameters in the \gls{pg} state, which starts at a temperature $\rm T_{mag}$ that  matches T$^*$ (see Figs. \ref{Fig:1} and  \ref{Fig:7}) and agrees with the downturn of the resistivity measurements in YBa$_2$Cu$_3$O$_{6+x}$ \cite{Fauque06}, in  HgBa$_2$CuO$_{4+\delta}$  \cite{Li08,Tang18} and in Bi$_2$Sr$_2$CaCu$_2$O$_{8+\delta}$  \cite{Watanabe97}. That suggested that the \gls{pg} line corresponds to a true phase transition. Later, detailed analysis of the magnetization measurements \cite{Leridon09}, resonant ultrasound spectroscopy \cite{Shekhter13} and torque 
measurements  \cite{Sato17,Murayama19} demonstrated in both YBa$_2$Cu$_3$O$_{6+x}$ and HgBa$_2$CuO$_{4+\delta}$  that the \gls{pg} line is truly a thermodynamic phase transition. 

Although the reported features at T$^*$ are rather weak in amplitude, they are all observed due to the higher accuracy of those techniques compared with specific heat data, where a large phonon background has to be removed. This establishes that the \gls{pg} state is necessarily associated with an ordered state, a point that should be present in any model relevant for the physics of  high-$\rm T_c$ cuprates.  For instance, it is argued \cite{Varma15}  that these results are compatible with the thermodynamics of the Ashkin–Teller model, which corresponds to the universality class of the \gls{lc}-ordered phase, as this model does not produce any strong jump in the specific heat for a wide range of parameters \cite{Gronsleth09}.

 It has to be stressed that no other true symmetry-breaking  phase than the \gls{iuc} magnetism develops  long range 3D correlation at  T$^*$. Contrariwise, the incipient  \gls{cdw} phase develops at lower temperature  (Fig. \ref{Fig:1}), deep inside the \gls{pg} state \cite{Keimer15,Proust19} and remains short range and quasi-2D.  Note that uniform perturbations, such as a strong external magnetic field\cite{Leboeuf13} or a uni-axial strain\cite{Kim18}, are required to promote  a long range ordered \gls{cdw}, at the expense of superconductivity. If uniform perturbations can be used to tune the balance between superconductivity and \gls{cdw}  inside the \gls{pg} state,  the substitution by a minute amount of non magnetic (Zn) impurity on Cu site is sufficient to  lower strongly the onset temperature of both states\cite{Blanco-Canosa13}. In contrast, the \gls{pg} state \cite{Alloul91} and  the onset temperature  of \gls{iuc} magnetism \cite{Baledent11} are both extremely robust against non magnetic impurity substitution.  Furthermore, the \gls{iuc} magnetism is reduced  in amplitude through Zn substitution \cite{Baledent11},  in agreement with a picture established by \gls{nmr} and \gls{stm} \cite{Alloul09} that the \gls{pg} physics basically vanishes near Zn impurities.

\subsection{Loop currents and pseudogap}


Because of its central play in the generic phase diagram, it is compulsory to consider \gls{pg} physics in any scenario aiming at a description of the physics of cuprates.  Being intimately tied to the \gls{pg} state, the \gls{iuc} magnetism, reported by \gls{pnd},  has to be implemented  in any theoretical approach. Although the interplay between the \glspl{lc} and the \gls{pg} state is rather obvious from experimental results, its theoretical importance remains unclear 
as  the proposed \gls{lc} order in cuprates cannot open a \gls{pg} at the Fermi level 
simply because it preserves  \gls{lt} symmetry\cite{Keimer15}. Indeed,  $\rm {\bf q}$=0 electronic instabilities in the electron-hole channel commonly 
split or distort the Fermi surface. This is the case for a ferromagnetic state, where two Fermi surfaces with different spin polarizations form in 
order to break the \gls{time}-symmetry, or for an electronic nematic state where the Fermi Surface  is distorted to break the \gls{rotation}- symmetry.  The only way to 
open a gap is then to break the \gls{lt}-symmetry.  Note, that superconductivity, which gaps the Fermi surface, is  also a $\rm {\bf q}$=0 electronic instability, but  in the electron-electron channel and it breaks a gauge symmetry.  Therefore, one is left with two possibilities that we discuss below: (i) there is an hidden modulation of the \gls{iuc} magnetism probed by \gls{pnd}, (ii) or the \gls{pg} state is a distinct state of matter, intertwined with the \gls{iuc} magnetism. 

A modified version of the \gls{lc} state, now breaking  \gls{lt}-symmetry, was recently proposed \cite{Varma19}. It consists of a \gls{lc} super-cell of the four degenerate \gls{lc} domains. Importantly,  the next-nearest domain of a given domain should be characterized by an anapole rotated by $\pi/2$, either turned left or right, whereas the domain with an anapole rotated by $\pi$ 
 is only joining the first domain through its corners. The domains size, counted in number of unit cells $P$, ranges from a few cells to dozens of 
them. Examples of such phases are depicted in Fig. \ref{Fig:2}.d for P=1 and in Fig. \ref{Fig:2}.e for P=2. It is shown that domains of $P\sim$ 10-20 unit cells give rise to both a \gls{pg} and Fermi arcs \cite{Varma19}, as experimentally evidenced by \gls{arpes} and \gls{stm} over the last two decades \cite{Keimer15}. In this approach, the new magnetic unit cell is square, $(2P \times 2 P)$ larger than the $\rm  CuO_2$ unit cell, and it is characterized by the planar wave vectors $ \rm {\bf q}_{\gls{lc}} $ =$(1/2P, 0)$ and $ (0, 1/2P)$. In a diffraction experiment like \gls{pnd} or resonant X-ray diffraction, this has major consequences, as the translation symmetry 
is now broken. Although this super-cell \gls{lc} state has not been observed experimentally yet, it cannot be excluded from the current \gls{pnd} data sets if 
$P$ is large enough. Indeed, the diffraction pattern has no maximum at the $(1,0,L)$ or $(0,1,L)$ Bragg reflections anymore, but gives rise to satellite reflections at $\rm  \pm {\bf q}_{\gls{lc}}$ away from them. If  $P$ is large enough, the diffracted peaks would occur as centered at the  Bragg positions owing to the broad neutron instrument resolution, and to the limitations of the experiments with respect to the finite neutron polarization\cite{Bourges11,Bourges18,Bourges19}.  Furthermore, if there were a distribution of domains size $P$ \cite{Varma14}, it would lead to an effective broadening of the magnetic peaks appearing on the shoulders of the crystal Bragg peaks.

 Keeping the initial uniform \gls{lc} phase, the latter can be considered as ancillary to a mother state associated with the \gls{pg} state. For instance, square lattice spin liquid states are predicted to possess  spontaneous \glspl{lc}  
\cite{Chatterjee17,Scheurer18}. In the framework of a fractionalized spin 
density wave, the \glspl{lc} also appear, forming an  ancillary order, resulting 
from the intertwining between a topological order and the discrete $Z_2$ broken symmetries.  In spin liquids,  the emergence of \gls{lc} orders in the archetype of doped 1D spin-liquids (hole-doped spin ladders) has been proposed \cite{Chudzinski08}.  Our recent results in spin-ladders \cite{Bounoua20} shows that local discrete symmetries are also broken in 1D spin liquid systems, as it has been  theoretically predicted. In other models, based on a \gls{pdw} state \cite{Agterberg15,Dai18} or a fractionalized \gls{pdw} instability\cite{Sarkar19}, the \gls{pg}  is described as a highly fluctuating state, where a preemptive phase breaking both \gls{time}- and \gls{parity}-symmetries (a \gls{lc}-like phase) is expected at higher temperature. More generally,  discrete symmetry breaking usually happens at higher temperature than continuous symmetry breaking. A strong influence of the \glspl{lc} state on both unidirectional and bidirectional $d$-wave charge-density-wave/pair-density-wave (\gls{cdw}/\gls{pdw}) composite orders has been emphasized \cite{Wang14,deCarvalho16}. In these approaches,  whatever the theoretical framework,  the \gls{iuc} \gls{lc} order has to be intertwined with other states 
of matter \cite{cai17}. For instance in ref.  \cite{Sarkar19}, one starts 
with two primary states reported at lower temperature deep in the \gls{pg} state (namely, the incipient \gls{cdw} and superconductivity, both carrying a $d$-wave symmetry). The \gls{pg} state is proposed to be a composite multi-component 
order parameter, where the system fluctuates between the two primary states which are intertwined. Then, the  \gls{lc} state would emerge  as  a preemptive state, born out of a higher order combination of the composite order parameters, which can be observed in the \gls{pg} state (as soon as it forms),  
whenever the primary states still fluctuate or become static on cooling down.

Another interesting question is the relationship of the \gls{iuc} magnetism with other $\bf q$=0 electronic instabilities. In addition to the electronic nematicity observed through the Nernst effect in YBa$_2$Cu$_3$O$_{6+x}$ effect at  T$^*$ \cite{Daou10}, enhanced electronic nematic fluctuations were observed close to the end point of the \gls{pg} state at large hole doping\cite{Auvray19}. However, they deviate from a canonical quantum critical scenario and cannot be directly associated with the \gls{pg} state.  Finally, 
the electronic nematicity could be inherited from another instability as suggested, for instance, by the sub-linear thermal dependence of torque measurements  \cite{Sato17,Murayama19}. This other instability could actually correspond to the \gls{lc} state.  Alternatively,  it could be a vestigial order \cite{Fradkin15} related to the \gls{cdw}, which exhibits a $d$-wave structure factor, implying that the electronic density on the O sites is involved \cite{Mukhopadhyay19}. At low temperature, the \gls{cdw} breaks both the \gls{lt} symmetry and the fourfold \gls{rotation}- symmetry. At high temperature, when the \gls{cdw} correlation length becomes  shorter than the  pitch of the modulation, the \gls{lt} symmetry is restored and,  the  electronic nematic correlation only is left inside the unit cell. Both states can be viewed as the nematic (\gls{rotation}- broken symmetry)  and smectic (\gls{lt}- broken symmetry) phases of an electronic liquid crystal. The role of \gls{iuc} magnetism in this last approach remains unclear. 

\subsection{\gls{iuc} time-scale} 

An important aspect of the \gls{iuc} magnetic order seen in \gls{pnd} is that the local magnetic probes like nuclear magnetic resonance (\gls{nmr})  and muon-spin rotation (\gls{musr}) are basically silent \cite{Bourges11} because detecting 
the corresponding magnetic fields with magnetic resonance techniques  has 
proved to be vastly unsuccessful, see {\it e.g.} in  HgBa$_2$CuO$_{4+\delta}$  \cite{Mounce13}.  Another attempt using \gls{nmr} in an ortho-II YBa$_2$Cu$_3$O$_{6+x}$ sample concluded that any static field at T=60K cannot be larger than $\sim$0.3 G  at apical O site  and $\sim $4 G at the planar 
O(2) sites\cite{Wu15}. For a static \gls{iuc} magnetic order, a magnetic field of one order of magnitude larger is expected on these sites. The fact that the existence of a static long ranged \gls{iuc} magnetism cannot be validated 
by \gls{nmr} is a serious issue, which questions the outcome of the \gls{pnd} measurements.  One possibility to reconcile \gls{nmr} and \gls{pnd} data is to consider internal fluctuating fields at a timescale slow enough to appear static at the timescale of \gls{pnd} scattering ($\sim$ 0.1 ns) but too fast to impact the \gls{nmr} lines ($\geq$ 10$\mu$s).

The situation with \gls{musr}  is more subtle. Since the first discovery of \gls{iuc} magnetism with \gls{pnd}, it has been envisaged \cite{Fauque06,Varma14,Pal16} that the magnetic moments could still fluctuate at the \gls{musr} timescale, meaning slowly enough to appear static to neutrons, but too fast to be observed as a magnetic order in \gls{musr}, therefore even more so for \gls{nmr}, which probes even slower timescales. Several controversial results were reported by one team \cite{Pal16,Pal18,Gheidi20} but as a matter of fact, this interesting picture has been experimentally approved, as slow magnetic fluctuations were discovered using a longitudinal field \gls{musr} technique in YBa$_2$Cu$_3$O$_{6+x}$ compounds  \cite{Zhang18,Zhu20} and in iridates \cite{Tan20}.

 In several YBa$_2$Cu$_3$O$_{6+x}$ underdoped samples  \cite{Zhang18,Zhu20},  at 80K just above the superconducting temperature,  the \gls{musr} measurements report magnetic correlations at a  finite timescale of $\sim$ 10 
ns which fluctuate slowly enough to give rise to resolution-limited magnetic peaks in a \gls{pnd} experiment. Furthermore, this time-scale does not appear to be strongly temperature dependent. Such a finite time-scale associated with finite spatial domains was proposed to happen if defects limit the magnetic correlation length \cite{Varma14}. Short range correlations with finite in-plane correlation lengths, $\xi \sim 20 a$, were actually detected in nearly optimally doped YBa$_2$Cu$_3$O$_{6+x}$  using \gls{pnd} \cite{Mangin15}. As discussed above,  shorter in-plane \gls{iuc} correlation lengths 
 are even observed in other cuprates, limited either by the competition with charge stripes in (La,Sr)$_2$CuO$_4$ \cite{Baledent10} or by dimension in two-leg spin ladder compounds \cite{Bounoua20}.  Clearly, dimension, disorder and  proliferation of magnetic domains associated with degenerate \gls{lc} states all can play a role. 

 The magnetic correlations seen by \gls{musr} can therefore be associated with slowly fluctuating magnetic domains. Interestingly, these fluctuations 
are maximum at the same temperature $\rm T_{mag}$ at which the neutron magnetic signal sets in \cite{Zhang18,Zhu20}. This suggests a critical slowing down of these magnetic fluctuations at  $\rm T^{*}$,  associated with 
the  \gls{pg} onset. Such dynamical fluctuations at  $\rm T^{*}$  have also been detected  by \gls{pnd}  in  nearly optimally doped  YBa$_2$Cu$_3$O$_{6+x}$  \cite{Mangin15} and in HgBa$_2$CuO$_{4+\delta}$  at two hole doping levels. \cite{Tang18}.  


\subsection{Possible roles of loop currents}

As it does not affect the \gls{lt} symmetry, the original \gls{lc} states are usually 
not considered as a  key player for the physics of  high-$\rm T_c$ superconducting  cuprates. It is typically much less discussed than \gls{cdw}, stripes physics or \gls{af} fluctuations \cite{Keimer15,Proust19}. This might be for 
two  main reasons. First, it is a more exotic phenomenon with, so far, a limited number of realizations in nature. The recent observation of  \gls{parity}-\cite{Zhao16}, \gls{time}-\cite{Jeong17} and  \gls{rotation}-\cite{Murayama20} symmetry breaking in (Sr,Rh)$_2$IrO$_4$ iridates clearly 
opens new frontiers as it basically 
corresponds to a \glspl{lc} picture\cite{Varma06,Chatterjee17}.  Second, it is a 
state which involves  not only copper orbitals but also oxygen orbitals. This mixing with an electronic phase running among the three $d$ and $p$ orbitals is a central feature, which imposes a three-band Hubbard model as a starting point \cite{Varma06,Weber14,Scheurer18}. A large majority of 
the models built for cuprates usually neglects the possible role of oxygen orbitals, as the band structure is nicely reproduced by a single renormalized band electronic structure observed in \gls{arpes} and \gls{stm} \cite{Keimer15}.

Numerical calculations using different methods still give contrasted pictures, even for three-band Hubbard model. On the one hand,  variational Monte Carlo calculations on asymptotically large lattices and exact diagonalization on a 24-site cluster shows a stabilization of the  \gls{lc}-$\Theta_{II}$  for a wide  range of parameters\cite{Weber14}. On the other hand, no 
instance of spontaneous \glspl{lc} has been found by applying cluster dynamical mean field theory  in ladder cuprates \cite{Lu20} in materials where they have been observed \cite{Bounoua20}. As a matter of fact,  variational Monte Carlo is a method to determine ground states at zero temperature, in contrast to the dynamical mean field theory, which is typically more efficient at finite temperature. 

Generally speaking, \glspl{lc} could be of great fundamental interest for several reasons. First, their fluctuations could be involved in the pairing mechanism and thus explain high-temperature superconductivity  \cite{Aji10}. 
Within that scenario, the electrons are paired through the coupling of their local angular momentum to an operator defined as the generator of the 
rotations in the space of the four degenerate states discussed above. The 
corresponding pairing vertex favors $d$-wave superconductivity and is also proportional to the \gls{lc} order susceptibility. Along similar lines, a novel spin-fluctuation-driven charge loop current mechanism based on the functional renormalization group theory has been proposed \cite{Tazai20}. The discussed mechanism leads to ferro-\gls{lc} order in a simple frustrated chain Hubbard model, which indicates that the \gls{lc} can be universal in strongly 
correlated electron metals near magnetic criticality with geometrical frustration.

 In the normal state, the related density susceptibility takes the functional form hypothesized for the Marginal Fermi liquid or strange metals \cite{Varma20}.  Namely, it should be rather independent of momentum and with almost structure-less energy dependence. Such a charge response has been recently measured in cuprates using a new technique probing the charge-charge correlation function \cite{Mitrano18}. It is argued that it corresponds to the density susceptibility expected for the \gls{lc} fluctuations in the normal state \cite{Varma20}. Interestingly, \gls{lc} states are thought to be an emergent phenomenon, important for the establishment of the strange 
metals physics \cite{Else20}.  More specifically,  if the scaling behavior of the conductivity is observed independent of impurities, they must be 
due to fluctuations of an order parameter which is a vector that breaks \gls{time} and  \gls{parity} symmetries and preserves their product \cite{Else20}. That is a precisely the statement for \gls{lc} order.
 
\gls{lc} orders could also be of particular importance in certain theories of the \gls{pg} state in cuprates in relation with topological orders \cite{Scheurer18} or emanating from the fluctuations of a vectorial order parameter \cite{Morice18,Sarkar19} as discussed above.

\section{\label{conclude} Conclusions and Perspectives}



In this brief overview, arguments were listed in favor of the physics of \glspl{lc} in quantum materials. Numerous experiments that include \gls{pnd} point towards the existence of an exotic \gls{lc} order parameter that does not break the \gls{lt}- symmetry. At a microscopic level, this corresponds to an electronic 
state phasing  $d$ orbitals of the transition metals with the ligand $p$ orbitals. To some extent, the observed phenomenon could also be described 
 by the ordering of magnetic quadrupoles of the transition metals \cite{Lovesey15,Fechner16}. All these order parameters, anapoles, quadrupoles or 
orbital currents curls have in common the breaking of both \gls{time}- and \gls{parity}- symmetries and	 the conservation of their product. These objects have dual (magnetic and electric) character order parameter, leading to interesting new applications for magneto-electricity \cite{Fiebig05}.  For instance, they are proposed for potential applications in the field of data storage, 
where it would be possible to control magnetization via the application of an electric field, and thus use these toroidal domains as magnetic qubit \cite{Spaldin08}.  

In this context, future developments will aim at exploring and generalizing the observation of these new exotic objects in a wider range of materials.   That will follow several  directions: for instance, the Fe-based superconducting materials could host as well  such a phase.  It is actually  proposed that the spin-orbit coupling present in the pnictides enforces the emergence of orbital loop current order inside the usual stripe-type spin-density wave state \cite{Klug18}.  In any case, the investigation of \glspl{lc} in other families of cuprate compounds using \gls{pnd}, namely  in CuO, Sr$_2$CuO$_2$Cl$_2$ and ${\rm La_{2-x}Ba_xCuO_4}$ should be pursued. In CuO, \glspl{lc}  were  reported by a resonant x-ray scattering study \cite{Scagnoli11}, though this interpretation is not necessarily supported by later experiments and simulations \cite{DiMatteo12,Joly12}. In  Sr$_2$CuO$_2$Cl$_2$, a  recent  optical \gls{shg} experiment points towards a magneto-chiral state \cite{delaTorre20}.  Following previous \gls{pnd} results \cite{Baledent10}, studying  ${\rm La_{2-x}Ba_xCuO_4}$ will allow one to the  study of  the competition between the \gls{iuc} magnetic order and other electronic instabilities, in that particular case the long range ordering of charge stripes 
\cite{Hucker11}. All these examples should address the important question of the coexistence of \gls{lc} phases with \gls{af} order in highly hole doped cuprates.


Another route to explore the magnetism related to the anapoles or quadrupoles is to use resonant X-ray diffraction at the appropriate atomic edge \cite{Scagnoli11,DiMatteo12,Joly12,Lovesey15b}. Unlike usual non-resonant 
X-ray diffraction, the  magnetic signal is enhanced at the resonant energy threshold of atomic electronic transitions. This technique,  sensitive to both  \gls{time}- and \gls{parity}- symmetries, allows the separation of the magnetic signal from the atomic signal, by the use of linear or circular polarization analysis of the scattered beam.  It allows one  to selectively probe dipole, quadrupole transitions or the interaction between these two processes. A key advantage is the ability to select the ion whose electronic transitions are induced by proper selection of the incident X-ray energy. For cuprates, measurements need to be carried at the absorption K-edge of Cu in order to access the terms corresponding to the anapoles, as derived from the multipolar expansion series. The \gls{iuc}  magnetic ordering observed 
in \gls{pnd} should induce an additional contribution to the anomal structure factor. For example, at the Cu K-edge (8985 eV), a dipolar-quadrupolar interference term (E1-E2) should occur with a characteristic dependence as a 
function of the azimuthal angle. This experiment certainly represents  an 
important challenge. 

Finally, the \gls{lc} story  has not been completed  yet. The exact interplay between \gls{lc} states and high-$\rm T_c$ superconductivity needs more experimental results, which will contribute to consolidate the theoretical models 
and to complete them. We hope our discovery will open up a new field of study when addressing the physical properties of highly correlated electron materials.

\section*{Acknowledgements}

The authors are deeply indebted to C. M. Varma who first proposed to search for  a loop currents phase in high-$\rm T_c$ cuprates and has been suggesting appealing novel ideas ever since. We   acknowledge also the many students and colleagues who contributed to this work either in France, V. 
Bal\'edent, S. de Almeida, B. Fauqu\'e, Jaehong Jeong, L.  Mangin-Thro, S.  Pailh\`es or within our international collaborations, M. Greven, Yuan Li and all our collaborators whose names appear in Ref.\cite{Fauque06,Mook08,Baledent11,Mangin15,Mangin17,Li08,Li11,Tang18,Baledent10,deAlmeida12,Mangin14,Jeong17,Bounoua20}. We wish to thank F. Damay, A.  Georges, T. 
Giamarchi, S.  Lovesey, C. P\'epin, E. Tailleur and  A. Shekhter for stimulating discussions on various aspects related to this work.  We acknowledge financial supports from the project NirvAna
(Contract ANR-14-OHRI-0010) of the Agence National de la Recherche (ANR)  French agency.

\printglossaries


\bibliographystyle{crunsrt}


 \bibliography{loop-revue}


\end{document}